# An Explainable Unsupervised-to-Supervised Machine Learning Framework for Dietary Pattern Discovery Using UK National Dietary Survey Data

Wing Yi Yu[1] and Chun Yin Chiu[2*]

1 School of Health Sciences, University of East Anglia, Norwich, United Kingdom; Wingyi.Yu@uea.ac.uk

2 Department of Informatics, King's College London, London, United Kingdom; *Corresponding author: Chun Yin Chiu, chun_yin.chiu@kcl.ac.uk

Abstract. Clinical dietary assessment can generate detailed but high-dimensional nutrient and food-group information that is difficult to translate quickly into counselling priorities. This paper proposes an explainable unsupervised-to-supervised machine learning framework for discovering, reproducing and interpreting dietary patterns using public UK National Diet and Nutrition Survey data. Adult participants aged 19 years and above from NDNS Years 12-15 were represented using 25 energy-adjusted nutrient and food-group features. K-means, Gaussian Mixture Models and Agglomerative Clustering were compared across k = 2-8, with stability and dietetic interpretability used alongside internal validation metrics. The selected K-means k = 4 solution identified four interpretable dietary patterns: high fat/meat and sodium, higher fibre fruit-vegetable micronutrient, high free-sugar snacks and sugary drinks, and dairy/cereal calcium-rich saturated-fat. A supervised surrogate classifier reproduced held-out cluster membership with high test performance (macro-F1 = 0.963), but was interpreted only as an explanatory surrogate rather than as an independent clinical prediction model. SHAP analysis linked predictions to dietetically meaningful drivers, suggesting potential value for dietitian-in-the-loop assessment, counselling prioritisation and follow-up monitoring.

## 1 Introduction

Dietary assessment is central to clinical dietetic practice, public health nutrition and nutritional epidemiology. Food diaries, 24-hour recalls and diet histories can produce detailed information on energy intake, macronutrients, micronutrients, sodium, fibre, sugars and food-group consumption. However, these variables are interrelated and high-dimensional. In clinical practice, dietitians must translate this information into practical counselling priorities, while in population nutrition researchers must summarise dietary behaviour without losing interpretability. Recent dietary assessment reviews continue to emphasise that measuring intake is challenging and subject to random and systematic error [1].

A further limitation of nutrient-by-nutrient interpretation is that individuals consume meals and combinations of foods rather than isolated nutrients. Dietary pattern analysis was developed as a complementary approach to single-nutrient analysis, and classic reviews established the value of empirical pattern discovery using factor or cluster analysis [2-3]. More recent methodological reviews show that dietary pattern research remains active, with growing use of machine learning, latent class analysis and other novel methods to capture dietary complexity [4].

Despite this progress, dietary pattern methods are not automatically useful for clinical dietetic practice. A cluster label alone does not tell a dietitian why an individual belongs to a pattern, which modifiable features should be discussed first, or how the output should be translated into patient-friendly advice. Recent reviews of AI in nutrition and clinical nutrition argue that AI tools should be clinically supervised, interpretable and carefully validated before being incorporated into care workflows [5-7]. This creates a need for dietary pattern models that are not only data-driven but also explainable and compatible with dietitian-in-the-loop practice.

This paper develops an explainable unsupervised-to-supervised machine learning framework for dietary pattern discovery using public dietary survey data. The framework first applies clustering to energy-adjusted dietary features, then trains supervised models to reproduce the derived cluster labels, and finally uses SHAP to identify the dietary features that distinguish the patterns. The supervised model is explicitly treated as a surrogate explanation model rather than an independent clinical prediction model.

The study makes three contributions. First, it proposes a reproducible dietary pattern discovery pipeline combining feature engineering, unsupervised clustering and supervised surrogate prediction. Second, it compares multiple clustering algorithms and classifier families using a public national dietary survey dataset. Third, it applies SHAP-based explainability to translate machine learning outputs into dietetically meaningful interpretations and discusses how such outputs could support clinical dietetic assessment, counselling prioritisation and monitoring. The overall framework is shown in Fig. 1.

### 1.1 Research Questions

The work was guided by three research questions. RQ1 asks whether unsupervised learning can identify interpretable dietary patterns from public dietary survey data. RQ2 asks which clustering configuration provides an acceptable balance between statistical quality, stability and dietetic interpretability. RQ3 asks whether supervised machine learning and explainability methods can reproduce and explain membership in AI-derived dietary pattern clusters.

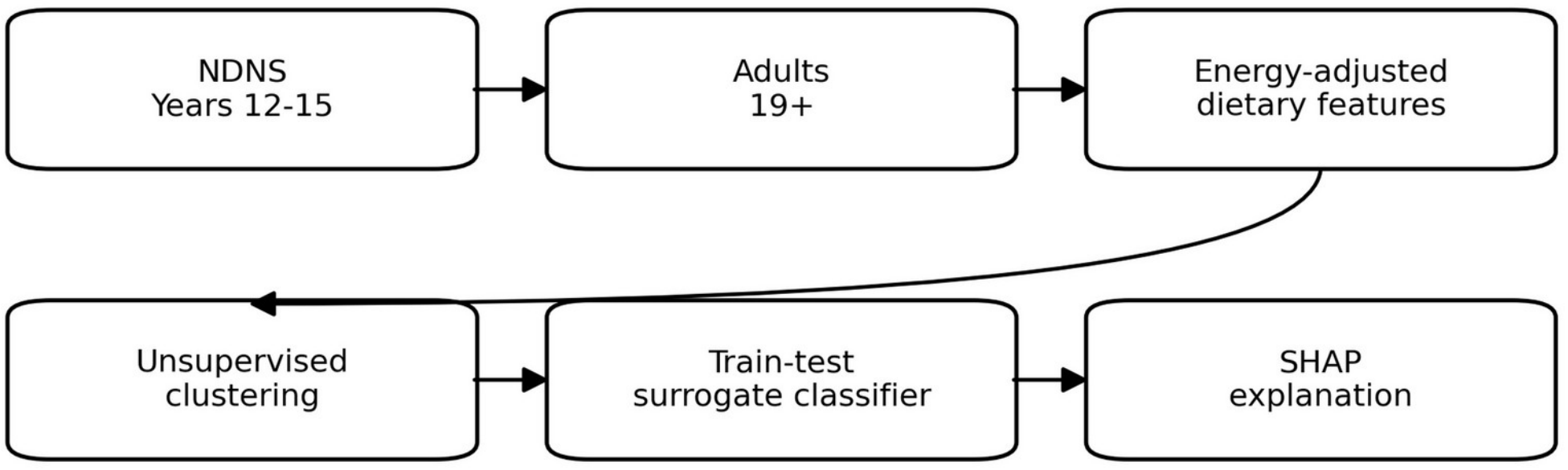

Fig. 1. Proposed unsupervised-to-supervised explainable dietary pattern discovery framework.

## 2 Related Work

Dietary pattern analysis has traditionally used predefined diet quality scores, principal component or factor methods, and cluster analysis. Predefined scores are interpretable but depend on external assumptions about healthy eating and nutrition assessment principles [8-10]. Data-driven approaches allow dietary patterns to emerge from observed intake data, but they require careful preprocessing because variables such as total energy intake, sodium or food-group quantities may dominate distance calculations if scales differ. Empirical dietary pattern research has been linked to health outcome interpretation as well as nutritional epidemiology more broadly [3, 11]. Recent reviews argue that novel methods, including machine learning and probabilistic approaches, may help characterise dietary complexity in greater depth than traditional approaches alone [4].

Clustering is widely used for exploratory pattern discovery. K-means provides a simple centroid-based baseline [12], Gaussian Mixture Models allow probabilistic assignments [13-14], and hierarchical methods can reveal nested structure [15]. Because different algorithms impose different assumptions, comparing multiple methods and cluster numbers is preferable to relying on a single run [16-17]. Internal validation measures such as silhouette score, Davies-Bouldin index and Calinski-Harabasz index provide quantitative evidence [18-20], although final selection also requires interpretability and practical cluster sizes.

Explainable machine learning is particularly important in applied health contexts. Feature-importance methods and SHAP values help identify which variables drive model decisions [21]. In nutrition and dietetics, this matters because the value of an AI-derived pattern depends not only on separability, but also on whether the pattern can be expressed in meaningful nutritional terms and reviewed by a professional. Recent AI nutrition reviews emphasise cautious integration as clinician-supervised decision support rather than autonomous decision-making [5-7].

The present work therefore integrates SHAP with a cluster-to-classifier workflow. The classifier is not presented as a disease-risk model; instead, it is a supervised surrogate used to test whether derived clusters can be reproduced on held-out data and to expose the dietary drivers behind cluster assignment.

## 3 Materials and Methods

### 3.1 Dataset and Study Population

The study used the UK National Diet and Nutrition Survey Rolling Programme, fieldwork Years 12-15, covering 2019 to 2023. NDNS is a continuous cross-sectional survey designed to assess diet, nutrient intake and nutritional status among people living in private households in the UK. For Years 12-15, dietary data were collected using Intake24, an online 24-hour dietary recall system, with up to four non-consecutive recalls per participant [22-23].

Two data files were used. The person-level dietary file contains mean nutrient intakes, food consumption variables, recipe food groups, disaggregated food variables and derived dietary variables. The individual-level file contains demographic, questionnaire, anthropometric and related survey variables. The files were merged using the individual serial identifier. The analytical sample was restricted to adults aged 19 years and above with available person-level dietary data, producing a final sample of 2,146 adults. The analytical sample and core dataset characteristics are summarised in Table 1.

Table 1. Analytical sample and core dataset characteristics.

| Characteristic | Value | Note |
|---|---|---|
| Dietary person-level records | 4,089 | Before adult restriction |
| Individual records | 4,370 | Used for age, sex and descriptive variables |
| Adult analytical sample | 2,146 | Participants aged 19+ with dietary data |
| Mean age | 52.0 years | Unweighted; weighted mean 49.5 years |
| Female participants | 55.3% | Unweighted; weighted estimate 51.9% |
| Mean energy intake | 1,626.5 kcal/day | Unweighted; weighted mean 1,647.4 kcal/day |
| Main clustering features | 25 | Energy-adjusted dietary features |

### 3.2 Dietary Feature Engineering

The unit of analysis was one participant represented by a participant-level dietary profile. To reduce the risk that clustering would be driven primarily by total energy intake, the main features were energy-adjusted. Macronutrient composition variables were expressed as percentage of total energy intake, while nutrient and food-group variables were transformed into densities per 1000 kcal.

The final feature set contained 25 dietary variables covering macronutrient composition, fat quality, sugar density, fibre density, sodium density, micronutrient density and selected food-group densities. Age, sex and other demographic variables were not included in the clustering feature set, ensuring that clusters represented dietary patterns rather than demographic profiles. The feature groups used in the main clustering analysis are listed in Table 2.

Table 2. Dietary feature groups used in the main clustering analysis.

| Feature group | Variables |
|---|---|
| Macronutrient composition | Protein, fat and carbohydrate percentage of energy |
| Fat quality | Saturated fat, monounsaturated fat and mono-to-saturated fat ratio |
| Carbohydrate quality | Total sugars and free sugars per 1000 kcal |
| Fibre and sodium density | AOAC fibre and sodium per 1000 kcal |
| Micronutrient density | Calcium, iron, folate, vitamin D and vitamin C per 1000 kcal |
| Food groups | Fruit, vegetables, fish, meat, milk, confectionery, sugary drinks, crisps/snacks, cereals and bread per 1000 kcal |

### 3.3 Preprocessing

NDNS missing value codes were treated as missing where appropriate. Continuous dietary features were winsorised at the 1st and 99th percentiles to reduce the influence of extreme values. Skewed variables were log-transformed where appropriate, remaining missing continuous values were imputed using median imputation, and all clustering variables were standardised using z-score scaling.

For the supervised stage, final cluster labels from the selected unsupervised solution were used as targets after preprocessing. An 80:20 stratified train-test split was used, and five-fold cross-validation was conducted on the labelled analytical sample using standard machine learning workflow components implemented in scikit-learn [24].

### 3.4 Unsupervised Dietary Pattern Discovery

Three clustering algorithms were evaluated: K-means, Gaussian Mixture Models and Agglomerative Clustering [12-15]. For each algorithm, candidate cluster numbers from $k = 2$ to $k = 8$ were tested. Cluster quality was evaluated using silhouette score, Davies-Bouldin index and Calinski-Harabasz index [18-20]. Cluster size distribution was examined to avoid solutions with very small or impractical groups.

Stability was assessed for shortlisted K-means solutions using repeated random seeds and bootstrap-based centroid reassignment, quantified by adjusted Rand index [25]. The final solution was selected by balancing statistical quality, stability, practical cluster size and dietetic interpretability. This is important because dietary intake data are noisy behavioural data, and the most interpretable solution is not always the one that maximises a single internal validity metric. To avoid overstating separability, the study reports the modest silhouette value directly and treats it as one component of evidence rather than as a pass-fail threshold.

### 3.5 Supervised Surrogate Prediction and SHAP Explainability

After clustering, the derived cluster labels were used as supervised targets. This design can appear circular if interpreted as ordinary outcome prediction, because the labels originate from the same dietary feature space. The supervised step was therefore not designed to predict an independent health outcome. Instead, it was used as a surrogate modelling and rule-extraction stage: the classifier evaluates whether the cluster assignments can be reproduced on held-out participants from the dietary feature representation and provides a model to which SHAP can be applied. Consequently, classifier performance is interpreted as cluster reproducibility and explainability evidence, not as evidence of clinical or external predictive validity.

Three classifiers were compared: Logistic Regression, Random Forest and XGBoost [26-27]. Performance was evaluated using accuracy, macro-F1 and weighted-F1, with macro-F1 treated as the primary metric because cluster sizes were not identical. SHAP analysis was applied to the best-performing classifier [21]. Global SHAP importance identified influential features overall, while class-specific SHAP summaries were used to interpret the dietary drivers of each pattern.

### 3.6 Sensitivity Analysis

Sensitivity analyses compared the main 25-feature energy-adjusted clustering solution with variants including total energy intake, no winsorisation, nutrient-only features and food-group-only features. Adjusted Rand index was used to measure concordance between each alternative solution and the main K-means $k = 4$ result.

## 4 Results

### 4.1 Cluster Model Selection

K-means generally produced more stable and interpretable clusters than Gaussian Mixture Models and Agglomerative Clustering. The K-means $k = 3$ solution had slightly stronger quantitative stability, but K-means $k = 4$ produced more dietetically differentiated patterns while maintaining practical cluster sizes.

The selected K-means $k = 4$ solution achieved a silhouette score of 0.0718, Davies-Bouldin index of 2.4822 and Calinski-Harabasz index of 200.30. Repeated-seed adjusted Rand index was 0.948 and bootstrap centroid adjusted Rand index was 0.676. Although the silhouette score was modest, cluster selection was not based on silhouette alone. In high-dimensional behavioural dietary data, low absolute silhouette values are expected when patterns overlap rather than form sharply separated natural classes. The four-cluster solution was preferred because it produced interpretable, non-trivial dietary profiles with the smallest cluster still representing 20.9% of the analytical sample. Table 3 summarises shortlisted solutions, and Fig. 2 compares silhouette scores across candidate clustering solutions. Sensitivity analysis also showed that simpler nutrient-only and food-group-only specifications gave higher silhouette values, but at the cost of losing the integrated nutrient-food interpretation that motivated the main framework.

Table 3. Summary of shortlisted clustering solutions.

| Alg. | k | Sil. | DB | CH | Seed ARI | Boot. ARI | Note |
|---|---|---|---|---|---|---|---|
| K-means | 3 | 0.090 | 2.470 | 233.47 | 0.983 | 0.786 | Broad |
| K-means | 4 | 0.072 | 2.482 | 200.30 | 0.948 | 0.676 | Selected |
| K-means | 5 | 0.069 | 2.512 | 184.64 | - | - | Frag. |

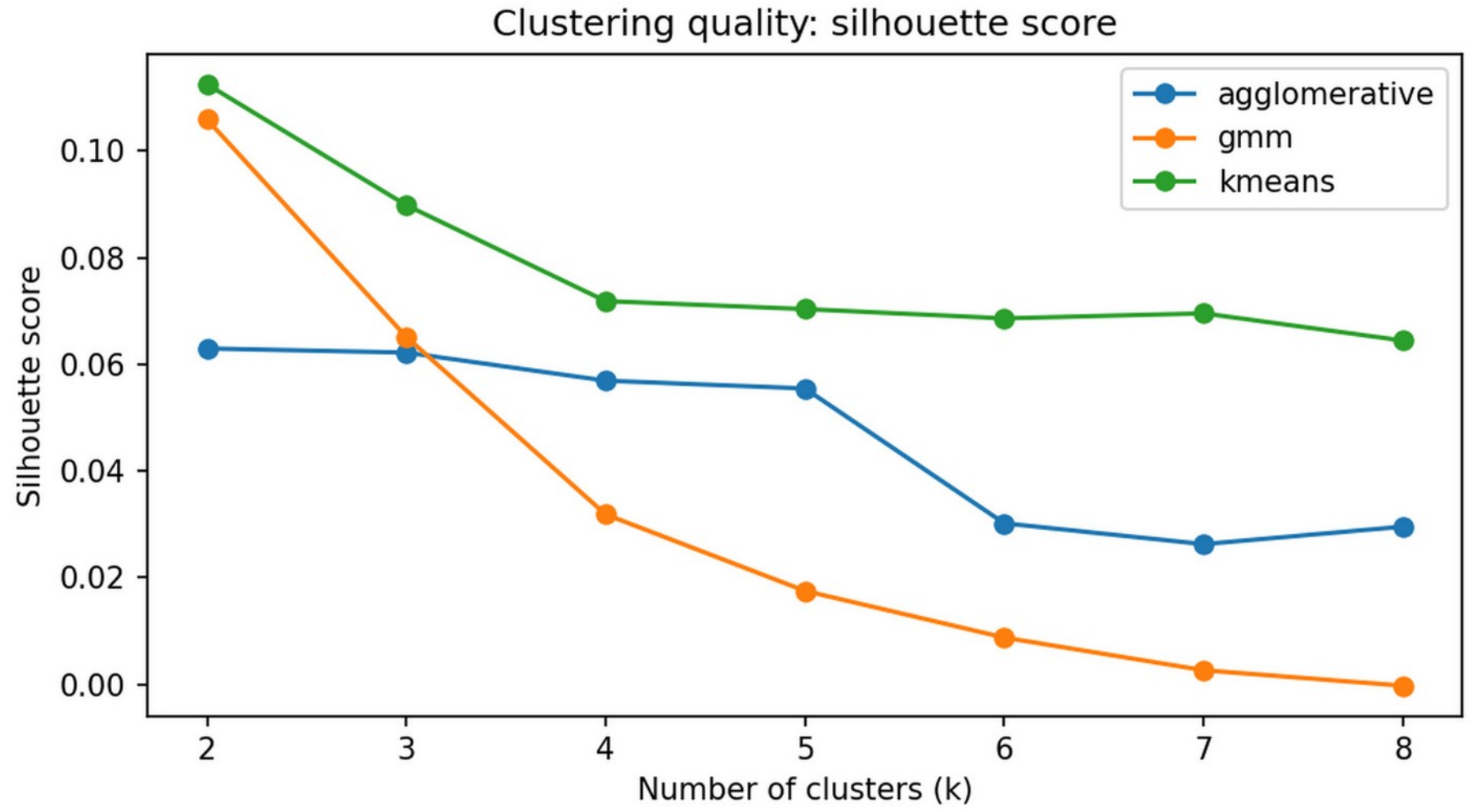

Fig. 2. Silhouette score comparison across candidate clustering solutions. Absolute scores were modest, so final selection also used stability, cluster size and dietetic interpretability.

### 4.2 Interpreted Dietary Patterns

The selected solution identified four dietary patterns with cluster sizes of 616, 449, 469 and 612 participants. Cluster names were assigned after examining standardised feature profiles rather than imposed a priori. The pattern interpretations are presented in Table 4, and their standardised feature profiles are visualised in Fig. 3.

Table 4. Dietary pattern interpretation for the selected K-means k = 4 solution.

| Cluster | Suggested name | n | % | Higher features | Lower features |
|---|---|---|---|---|---|
| 0 | High fat/meat and sodium pattern | 616 | 28.7% | monounsaturated fat g/1000 kcal; meat g/1000 kcal; fat pct energy; mono to saturated fat ratio | total sugars g/1000 kcal; carbohydrate pct energy; free sugars g/1000 kcal; milk g/1000 kcal |
| 1 | Higher fibre fruit-vegetable micronutrient pattern | 449 | 20.9% | fibre g/1000 kcal; folate ug/1000 kcal; fruit g/1000 kcal; iron mg/1000 kcal | saturated fat g/1000 kcal; fat pct energy; monounsaturated fat g/1000 kcal; meat g/1000 kcal |
| 2 | High free-sugar snacks and sugary drinks pattern | 469 | 21.9% | free sugars g/1000 kcal; sugary drinks g/1000 kcal; total sugars g/1000 kcal; confectionery g/1000 kcal | protein pct energy; folate ug/1000 kcal; fibre g/1000 kcal; iron mg/1000 kcal |
| 3 | Dairy/cereal calcium-rich saturated-fat pattern | 612 | 28.5% | calcium mg/1000 kcal; milk g/1000 kcal; saturated fat g/1000 kcal; breakfast cereals g/1000 kcal | mono to saturated fat ratio; crisps savoury snacks g/1000 kcal; sugary drinks g/1000 kcal; vegetables g/1000 kcal |

Cluster 0 represented a high fat/meat and sodium pattern. It had higher monounsaturated fat density, meat intake, fat percentage, monounsaturated-to-saturated fat ratio, sodium density and protein percentage, with lower sugar-related and milk/calcium features.

Cluster 1 represented a higher fibre fruit-vegetable micronutrient pattern. It had higher fibre, folate, fruit, iron, vitamin C and vegetable density, and lower saturated fat, total fat, monounsaturated fat, meat, sugary drinks and savoury snacks. This was the most favourable nutrient-density profile.

Cluster 2 represented a high free-sugar snacks and sugary drinks pattern. It had higher free sugars, sugary drinks, total sugars, confectionery, carbohydrate percentage and crisps or savoury snacks, with lower protein, folate, fibre, iron, calcium and vegetable intake.

Cluster 3 represented a dairy/cereal calcium-rich saturated-fat pattern. It had higher calcium, milk, saturated fat and breakfast cereal intake, but lower monounsaturated-to-saturated fat ratio, savoury snacks, sugary drinks, vegetables, meat and vitamin C.

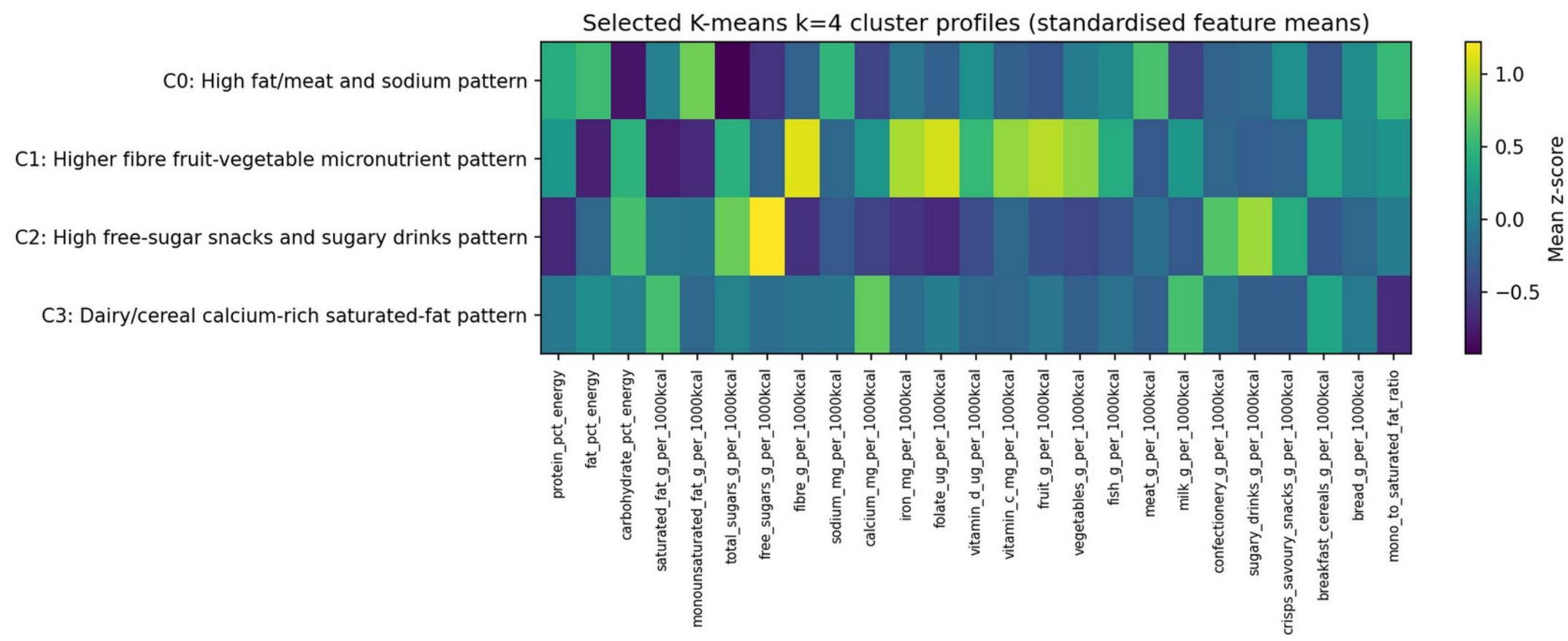

Fig. 3. Standardised feature profiles for the selected four dietary patterns.

### 4.3 Supervised Surrogate Classifier Performance

The derived cluster labels were used as supervised targets. Logistic Regression achieved the strongest held-out test performance, with accuracy = 0.963, macro-F1 = 0.963 and weighted-F1 = 0.963. XGBoost achieved macro-F1 = 0.920 and Random Forest achieved macro-F1 = 0.887. Table 5 reports the full supervised surrogate performance comparison.

Five-fold cross-validation confirmed this result. Logistic Regression obtained mean macro-F1 = 0.955 +/- 0.007, compared with 0.905 +/- 0.008 for XGBoost and 0.884 +/- 0.011 for Random Forest. This suggests that the cluster structure was highly reproducible and largely separable by relatively linear combinations of the energy-adjusted dietary features. Fig. 4 compares macro-F1 across classifiers, and Fig. 5 shows the held-out confusion matrix for the best model.

Table 5. Supervised surrogate classifier performance for predicting cluster membership.

| Model | Accuracy | Macro-F1 | Weighted-F1 | 5-fold CV macro-F1 |
|---|---|---|---|---|
| Logistic Regression | 0.963 | 0.963 | 0.963 | 0.955 +/- 0.007 |
| XGBoost | 0.921 | 0.920 | 0.921 | 0.905 +/- 0.008 |
| Random Forest | 0.888 | 0.887 | 0.888 | 0.884 +/- 0.011 |

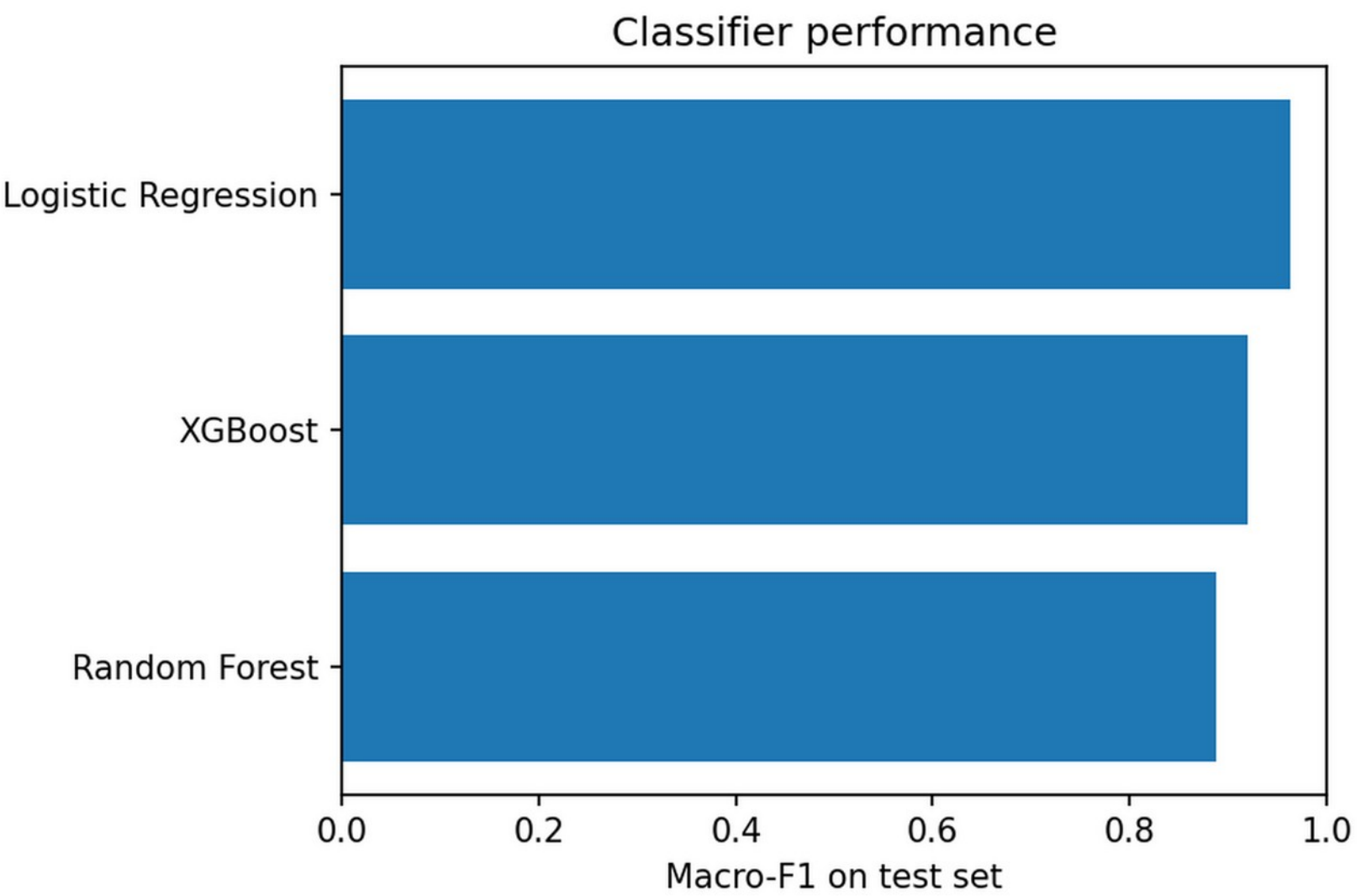

Fig. 4. Macro-F1 comparison across supervised surrogate classifiers.

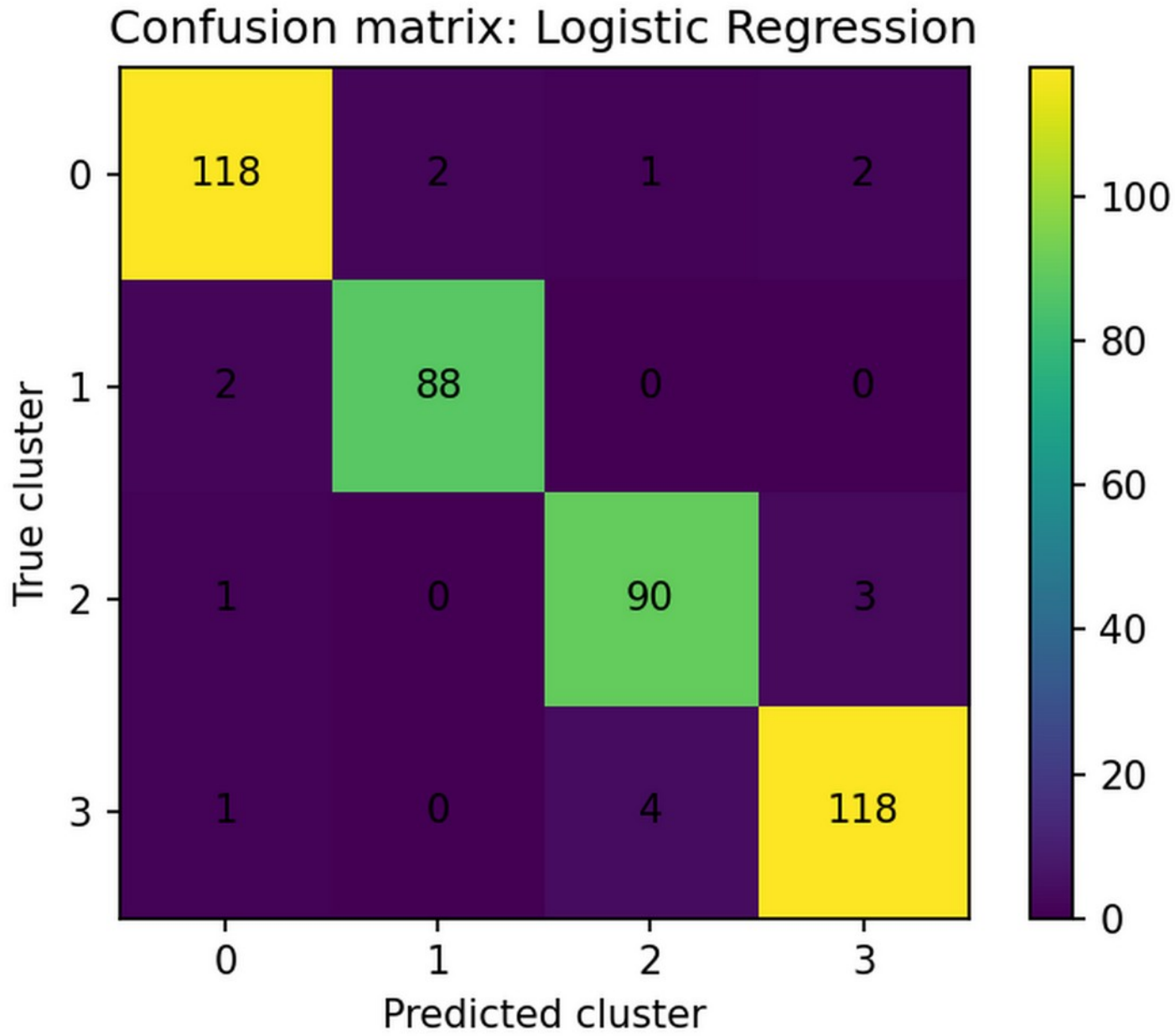

Fig. 5. Confusion matrix for the best-performing supervised surrogate classifier on the held-out test set.

### 4.4 Explainability Analysis

Global SHAP analysis indicated that cluster prediction was primarily driven by total sugars, folate, iron, free sugars, fibre, fruit, milk, monounsaturated fat, vegetables, carbohydrate percentage, meat, saturated fat and calcium. These variables correspond closely to the dietary interpretation of the four clusters. Table 6 and Fig. 6 show the global SHAP importance results.

Table 6. Top global SHAP feature importances for the best classifier.

| Feature | Mean absolute SHAP value |
|---|---|
| total sugars g/1000 kcal | 0.996 |
| folate ug/1000 kcal | 0.911 |
| iron mg/1000 kcal | 0.839 |
| free sugars g/1000 kcal | 0.822 |
| fibre g/1000 kcal | 0.818 |
| fruit g/1000 kcal | 0.797 |
| milk g/1000 kcal | 0.786 |
| monounsaturated fat g/1000 kcal | 0.697 |
| vegetables g/1000 kcal | 0.672 |
| carbohydrate pct energy | 0.653 |
| meat g/1000 kcal | 0.643 |
| saturated fat g/1000 kcal | 0.639 |

The SHAP results supported the domain plausibility of the clusters. Sugar-related variables helped distinguish the high free-sugar snacks and sugary drinks pattern. Fibre, fruit, vegetables, folate, iron and vitamin C helped distinguish the higher fibre fruit-vegetable micronutrient pattern. Milk and calcium contributed to the dairy/cereal calcium-rich pattern, while fat, meat, sodium and protein-related variables contributed to the high fat/meat and sodium pattern. Table 7 summarises the class-specific SHAP drivers, and Table 8 gives local participant-level examples.

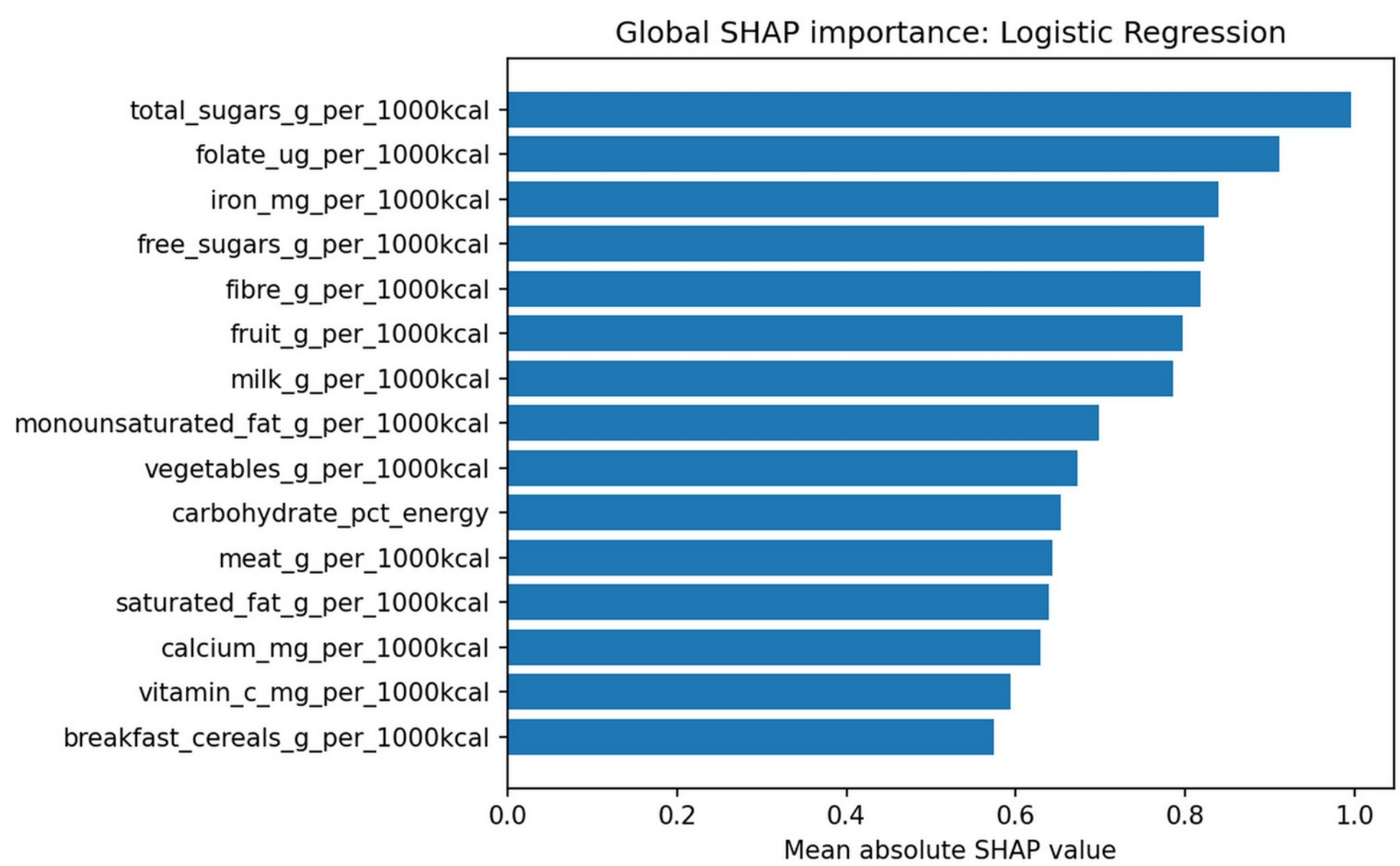

Fig. 6. Global SHAP feature importance for the best supervised surrogate classifier.

Table 7. Class-specific SHAP drivers by dietary pattern.

| Cluster | Pattern | Top class-specific SHAP drivers |
|---|---|---|
| 0 | High fat/meat and sodium pattern | total sugars g/1000 kcal; monounsaturated fat g/1000 kcal; carbohydrate pct energy; meat g/1000 kcal; sodium mg/1000 kcal |
| 1 | Higher fibre fruit-vegetable micronutrient pattern | folate ug/1000 kcal; iron mg/1000 kcal; fibre g/1000 kcal; fruit g/1000 kcal; vegetables g/1000 kcal |
| 2 | High free-sugar snacks and sugary drinks pattern | free sugars g/1000 kcal; total sugars g/1000 kcal; iron mg/1000 kcal; protein pct energy; sugary drinks g/1000 kcal |
| 3 | Dairy/cereal calcium-rich saturated-fat pattern | milk g/1000 kcal; saturated fat g/1000 kcal; calcium mg/1000 kcal; mono to saturated fat ratio; meat g/1000 kcal |

Table 8. Local SHAP examples showing participant-level positive drivers.

| Predicted cluster | Example ID | Top positive local drivers |
|---|---|---|
| 0 | 150182242 | carbohydrate pct energy; monounsaturated fat g/1000 kcal; meat g/1000 kcal |
| 1 | 140041181 | iron mg/1000 kcal; folate ug/1000 kcal; fibre g/1000 kcal |
| 2 | 140030191 | free sugars g/1000 kcal; sugary drinks g/1000 kcal; total sugars g/1000 kcal |
| 3 | 140211341 | calcium mg/1000 kcal; milk g/1000 kcal; mono to saturated fat ratio |

### 4.5 Sensitivity Analysis

The main energy-adjusted feature set was compared with several alternative specifications. Adding total energy intake produced a highly similar solution, with adjusted Rand index of 0.921 relative to the main K-means k = 4 clustering. This indicates that the selected clusters were not simply driven by total energy intake.

Removing winsorisation produced a moderately similar solution with adjusted Rand index of 0.768, suggesting that outlier handling influenced but did not determine the clustering. Nutrient-only and food-group-only feature sets produced more different results, with adjusted Rand indices of 0.438 and 0.244 respectively. Table 9 reports the sensitivity results, supporting the use of a combined nutrient and food-group representation for dietary pattern discovery.

Table 9. Sensitivity analysis relative to the main K-means k = 4 solution.

| Analysis | Features | Sil. | ARI vs main | Cluster sizes |
|---|---|---|---|---|
| full energy adjusted features | 25 | 0.072 | 1.000 | 616;449;469;612 |
| full plus total energy | 26 | 0.070 | 0.921 | 632;625;460;429 |
| full no winsorisation | 25 | 0.069 | 0.768 | 610;492;561;483 |
| nutrient only features | 15 | 0.110 | 0.438 | 424;771;466;485 |
| food group only features | 10 | 0.102 | 0.244 | 646;600;405;495 |

## 5 Discussion

The results show that an explainable unsupervised-to-supervised framework can identify meaningful dietary patterns from public dietary survey data. The final four-cluster solution separated adults into interpretable profiles corresponding to higher fruit-vegetable micronutrient density, high sugar and snack intake, higher fat/meat and sodium intake, and a dairy/cereal calcium-rich profile.

The study also demonstrates why cluster evaluation in applied health and dietetics should not rely on a single numeric criterion. The silhouette score of the selected solution was low in absolute terms, and this is acknowledged as a limitation rather than hidden. However, dietary behaviour rarely forms sharply separated natural classes; individuals often sit on continua of sugar intake, fibre density, food-group composition and fat quality. The selected solution was supported by practical cluster sizes, stability checks, sensitivity analysis and dietetic interpretability. In addition, the sensitivity analyses showed the expected trade-off: narrower feature sets improved silhouette modestly, while the combined feature set produced richer nutritional interpretation. Therefore, the contribution is not the discovery of perfectly separated dietary classes, but a transparent framework for deriving and explaining useful dietary profiles.

A major methodological issue is the apparent circularity of predicting cluster labels using features similar to those used to generate the labels. This study addresses that concern in three ways. First, the paper explicitly defines the classifier as a surrogate modelling and explanation stage rather than as independent outcome prediction. Second, performance is reported using held-out supervised evaluation and is interpreted as reproducibility of the derived dietary pattern rules, not as external validity. Third, SHAP is used to explain the surrogate decision function in nutritional terms, not to claim causality. The classifier therefore asks whether the discovered clusters can be reproduced and explained, not whether disease, health status or objectively true dietary categories can be predicted.

The high performance of Logistic Regression is informative. In many applied machine learning studies, more complex models outperform linear baselines. Here, the linear model outperformed Random Forest and XGBoost, suggesting that the selected clusters were separated by relatively linear combinations of standardised dietary features. This strengthens interpretability and reduces reliance on complex black-box modelling.

The SHAP analysis further supports the plausibility of the framework. The most important features were dietetically meaningful rather than arbitrary: sugar density, fibre, fruit, vegetables, micronutrients, milk, calcium, fat, meat and sodium. These features map directly onto the interpreted cluster labels and provide a bridge between machine learning outputs and nutritional interpretation.

## 6 Practical and Clinical Dietetic Implications

The main practical value of the framework is not automated diagnosis, but AI-assisted dietary pattern recognition. Current dietetic assessment often requires a dietitian to inspect food diaries, 24-hour recall outputs or nutrient reports and manually integrate many signals, including sugars, fibre, sodium, saturated fat, fruit, vegetables and micronutrient density. This professional judgement is essential, but the process can be time-consuming and may make it difficult to see recurring combinations of dietary behaviours. The proposed framework summarises complex intake data into interpretable dietary profiles while still showing which features drive the assignment.

In a clinical workflow, a patient could complete a 24-hour recall or food diary that is converted into the same type of energy-adjusted dietary features used in this study. The system would then assign the patient to the nearest dietary pattern and display the leading SHAP drivers. For example, a patient assigned to the high free-sugar snacks and sugary drinks pattern may be driven by high free sugars, sugary drinks, confectionery and low fibre. Rather than giving generic advice to eat more healthily, the dietitian could prioritise sugary drink replacement, snack reformulation and fibre-rich alternatives. A patient assigned to the high fat/meat and sodium pattern may instead receive counselling focused on processed meat, salt reduction, fat quality and portion balance.

This pattern-level output may also improve communication with patients. Nutrient tables are often difficult to translate into plain language, whereas a pattern explanation can be phrased around behaviours: sugary drinks, sweet snacks, low fibre, low fruit and vegetable intake, processed meat, sodium or saturated fat. SHAP explanations can support shared goal-setting by identifying one or two modifiable drivers to address first rather than overwhelming the patient with a full nutrient report.

The framework could also support follow-up monitoring. At a subsequent appointment, updated dietary recall data could be processed to assess whether the patient remains in the same pattern, shifts toward a higher fibre fruit-vegetable micronutrient pattern, or shows reduced SHAP contribution from a target driver such as sugary drinks or sodium. Even without a full cluster change, a reduction in a problematic driver may provide meaningful feedback for counselling and motivation.

At service level, clinics could use aggregated pattern distributions to plan targeted education. A diabetes prevention, weight management or cardiovascular risk service may discover that many patients fall into a high sugar/snack profile, while another clinic may see more high sodium/meat profiles. Group education can then be designed around common dietary patterns rather than generic nutrition messages. The system should remain dietitian-in-the-loop: clusters are exploratory dietary profiles, not diagnoses, and outputs should be interpreted alongside clinical history, biochemical data, anthropometry, medication, cultural preferences, socioeconomic context and patient goals.

## 7 Limitations

This study has several limitations. First, dietary recall data may contain self-reporting error and day-to-day variation. The person-level NDNS variables reduce within-person variation through usual-intake or day-average estimates, but measurement error cannot be eliminated. Second, the clustering labels are algorithm-derived and should not be interpreted as objective clinical categories.

Third, internal clustering metrics such as silhouette and Davies-Bouldin scores were modest, so the clusters should be treated as useful dietary profiles rather than sharply separated natural classes. Fourth, survey weights were not directly incorporated into the machine learning algorithms. Weighted descriptive summaries may be used for population reporting, but the main purpose here was pattern discovery rather than formal prevalence estimation.

Finally, the analysis used NDNS Years 12-15 only. This avoids mixing different dietary assessment methods across NDNS phases, but it also means that the results should be interpreted as patterns in the 2019-2023 Intake24 data rather than a full historical analysis of Years 1-15. Future work could evaluate whether similar patterns appear in other survey waves or datasets.

## 8 Reproducibility

All transformations were defined as deterministic preprocessing steps before clustering. The workflow stores the adult analytical dataset, feature transformations, clustering metrics, cluster profiles, classifier performance tables and SHAP summaries as separate output files. This supports reproducibility and allows the reported results to be regenerated without manual spreadsheet editing.

The analysis also reports both favourable and unfavourable evidence. The selected clustering solution is interpretable and reproducible, while its modest silhouette score and surrogate classifier design are explicitly acknowledged. This reduces overclaiming and clarifies that the framework is intended for exploratory dietary pattern discovery and decision support rather than autonomous clinical decision-making.

## 9 Conclusion

This paper presented an explainable unsupervised-to-supervised machine learning framework for dietary pattern discovery using public UK dietary survey data. Using 2,146 adults from NDNS Years 12-15 and 25 energy-adjusted dietary features, the framework identified four interpretable dietary patterns. A supervised surrogate classifier reproduced cluster membership with high performance, and SHAP analysis linked model predictions to meaningful nutritional drivers.

The findings suggest that combining clustering, supervised surrogate prediction and explainable AI can provide a useful methodological bridge between machine learning and dietetic interpretation. Future work could extend the framework to children and adolescents, incorporate other survey waves, and investigate associations between AI-derived dietary patterns and health outcomes.

### Data Availability

The data used in this study are available from the UK Data Service as part of the National Diet and Nutrition Survey Rolling Programme, subject to UK Data Service access conditions.

### Acknowledgements

The authors acknowledge the UK Data Service and the NDNS research teams for making the survey data and documentation available for research use.